\documentclass[sn-mathphys,Numbered]{sn-jnl}%

\usepackage{graphicx}%
\usepackage{multirow}%
\usepackage{amsmath,amssymb,amsfonts}%
\usepackage{amsthm}%
\usepackage{mathrsfs}%
\usepackage[title]{appendix}%
\usepackage{xcolor}%
\usepackage{textcomp}%
\usepackage{manyfoot}%
\usepackage{booktabs}%
\usepackage{algorithm}%
\usepackage{listings}%
\usepackage{times}
\usepackage{epsfig}
\usepackage{physics} 
\usepackage{bbm} 

\usepackage{comment}
\usepackage{algorithmic}
\usepackage{algorithm}
\usepackage{nicefrac}
\usepackage{caption,tabularx,booktabs}
\usepackage{caption}
\usepackage{subcaption}

\newcommand{\Rone}[1]{\textcolor{black}{#1}}

\newcommand{\RTwoTwo}[1]{\textcolor{black}{#1}}

\newcommand{\RTwoThree}[1]{\textcolor{black}{#1}}

\theoremstyle{thmstyleone}%

\theoremstyle{thmstyletwo}%

\theoremstyle{thmstylethree}%

\begin{document}

\title[Article Title]{Quantum Vision Clustering}

\author[1, 3]{\fnm{Xuan Bac} \sur{Nguyen}}\email{xnguyen@uark.edu}

\author[2, 3]{\fnm{Hugh} \sur{Churchill}}\email{hchurch@uark.edu}

\author*[1, 3]{\fnm{Khoa} \sur{Luu}}\email{khoaluu@uark.edu}

\author[4]{\fnm{Samee} \sur{U. Khan}}\email{skhan@ece.msstate.edu}

\affil[1]{\orgdiv{CVIU Lab, Department of Electrical Engineering and Computer Science}, \orgname{University of Arkansas}, \orgaddress{\city{Fayetteville}, \postcode{72701}, \state{Arkansas}, \country{USA}}}

\affil[2]{\orgdiv{Department of Physics}, \orgname{University of Arkansas}, \orgaddress{\city{Fayetteville}, \postcode{72701}, \state{Arkansas}, \country{USA}}}

\affil[3]{\orgdiv{MonArk NSF Quantum Foundry}, \orgname{University of Arkansas}, \orgaddress{\city{Fayetteville}}}

\affil[4]{\orgdiv{Department of Electrical and Computer Engineering}, \orgname{Mississippi State University}, \orgaddress{\city{Starkville}, \postcode{39762}, \state{Mississippi}, \country{USA}}}

\abstract{
Unsupervised visual clustering has garnered significant attention recently, aiming to characterize distributions of unlabeled visual images through clustering based on a parameterized appearance approach. Alternatively, clustering algorithms can be viewed as assignment problems where an unlabelled sample needs to be assigned to a specific cluster. This problem \RTwoTwo{can be formulated as Quadratic Unconstrained Binary Optimization that is} characterized as NP-hard, yet precisely solvable for small instances on contemporary hardware. \Rone{Adiabatic quantum computing (AQC) emerges as a promising solution to offer a good speed in solving the NP-hard optimization problems}. \Rone{However, existing clustering approaches face challenges in designing the problems to be solved by the quantum computer}. In this study, we present the first clustering formulation tailored for resolution using Adiabatic quantum computing. An Ising model is introduced to represent the quantum mechanical system implemented on AQC. The proposed approach demonstrates high competitiveness compared to state-of-the-art optimization-based methods, even when utilizing off-the-shelf integer programming solvers. Lastly, this work showcases the solvability of the proposed clustering problem on current-generation real quantum computers for small examples and analyzes the properties of the obtained solutions.}


\maketitle

\section*{Article Highlights}
\begin{itemize}
    \item The paper presents a novel Quadratic Unconstrained Binary Optimization (QUBO) formulation for unsupervised clustering.
    \item The paper leverages quantum mechanics to solve the QUBO problem.
    \item The empirical experiments demonstrate the efficientness of the quantum computers in solving unsupervised clustering problems.
\end{itemize}

\section{Introduction}
Unsupervised learning in automated object and human understanding has recently become one of the most popular research topics in robot vision. It is primarily due to the abundance of unlabelled raw data and the need for robust visual recognition algorithms that can perform consistently in various challenging conditions. Traditional methods, like K-Mean clustering, have shown limitations regarding speed and accuracy when dealing with large-scale databases. These methods heavily rely on a predetermined number of clusters ($k$), which is not applicable in real scenarios, such as a visual object or visual landmark recognition clustering. In recent years, several studies \cite{yang2019learning,yang2020learning, Nguyen_2021_CVPR} based on deep learning have been proposed to address the challenge of unsupervised clustering. However, these methods rely on rule-based algorithms for the final cluster construction. Meaning they are still required to take additional steps to process large-scale databases.

In this paper, we will map the clustering problem into a quantum mechanical system whose energy is equivalent to the cost of the optimization problem. Therefore, if it is possible to measure the lowest energy state of the system, a solution to the corresponding optimization problem can be derived. It is done with an Adiabatic Quantum Computer (AQC), which implements a quantum mechanical system made from qubits and can be described by the Ising model. Using this approach, a quantum speedup, which further scales with system size and temperature, has already been shown for applications in physics.

While quantum computing can provide a range of future advantages, \Rone{formulating a problem in the form of AQC is a nontrivial task}. It often requires reformulating the problem from scratch, even for well-investigated tasks. Since the problem needs to be matched to the Ising model, real quantum computers have a minimal number of qubits and are still prone to noise, which requires tuning the model to handle the limitations. In this work, we present the first quantum computing approach for visual clustering.

\textbf{Principal Contributions}: In this work, we introduce a novel visual clustering approach using quantum machine learning. Our principal contributions can be summarized as follows: 
\begin{itemize}
    \item Initially, we formulate the clustering problem as a Quadratic Unconstrained Binary Optimization (QUBO) problem. Unlike previous methods, our formulation allows us not to need to know the number of clusters.
    \item We present an efficient solution approach for the QUBO problem using Adiabatic Quantum Computing, aiming to attain promising unsupervised clustering outcomes.
    \item The results from the experiment illustrate the feasibility of our proposed solution for the clustering problem. 
\end{itemize}

Throughout the rest of this paper, we will present related studies addressing the visual clustering problem. Subsequently, we articulate the visual clustering problem within the quantum framework. It will be followed by an exploration of fundamental concepts in quantum computing. We will then detail our approach to framing the visual clustering problem through a QUBO formulation. Lastly, we demonstrate the outcomes of our experiments conducted on the D-Wave quantum computer and draw comparisons with the $k$-NN clustering algorithm.

\section{Related Work}
\textbf{Quantum Machine Learning}: Quantum machine learning algorithms have been recently developed to solve problems that are too slow to solve on classical computers \cite{kerenidis2019q, cerezo2022challenges, nguyen2024theory, schatzki2024theoretical}
A class of discrete optimization problems known as Quadratic Unconstrained Binary Optimization problems is particularly amenable to a technique known as Adiabatic Quantum computing. This paper investigates how classical computing and Adiabatic Quantum computing can be unified to solve the Visual Clustering problem by formulating it in an end-to-end QUBO framework. Prior studies have used the strategy of solving QUBO problems with Adiabatic Quantum computing for tasks like cluster assignment, robust fitting, and $k$-means clustering. \cite{zaech2022adiabatic, doan2022hybrid, arthur2021balanced}
\newline
\noindent
\textbf{Deep Visual Clustering}: Visual clustering is a crucial tool for data analytics. Deep visual clustering refers to applying deep learning methods to clustering. Classical techniques for clustering such as density-based clustering, centroid-based clustering, distribution-based clustering, ensemble clustering, etc.  \cite{ren2022deep} display limited performance on complex datasets. Deep neural networks seek to map complex data to a feature space where it becomes easier to cluster. \cite{ren2022deep} Modern deep clustering methods have focused on Graph Convolutional Neural Networks (GCN) and Transformers \cite{Lopes_2023_WACV, bo2020structural, ling2022vision}. Empirical data is often most naturally represented with a graph structure. For example, websites are usually connected by a hyperlink. Social media platform users are connected by being followers or stores connected by roads. 
GCNs exploit this graph structure by taking it as input to produce more accurate features, which are used finally to cluster the nodes. \cite{chen2022graph, kipf2016semi, huo2021caegcn, 4700287} Transformers also exploit the connections between the data; Ling et al. unified for the first time the transformer model and contrastive learning for image clustering. \cite{ling2022vision} Nguyen et al. presented the Clusformer model, which addresses the sensitivity of the GCN to noise and uses a transformer for the task of visual clustering \cite{nguyen2021clusformer}.

\section{Preliminaries on Quantum Computing}

\subsection{Quantum Bit (Qubit)} 
A qubit in a quantum computer has the role as the analog of a bit in a classical computer. Mathematically, a qubit $q$ is a vector in a two-dimensional complex vector space spanned by the basis vectors $\ket{0} = [1, 0]^T$ and $\ket{1} = [0, 1]^T$: 
\begin{equation*}
    q = \alpha \ket{0} + \beta \ket{1} 
\end{equation*}
such that $|\alpha|^2 + |\beta|^2 = 1$. Physically, a qubit can be realized as the different energy levels of an atom, polarizations of a photon, ions trapped in an electromagnetic stasis, and more. \cite{rieffel2011quantum}

\subsection{\RTwoTwo{Quantum Superposition}}
\RTwoTwo{A qubit is in quantum superposition if its state cannot be expressed as a single-scaled basis vector \cite{rieffel2011quantum}. For example, qubit $q_1$ is in quantum superposition and qubit $q_2$ is not in the following equations}:
\begin{equation*}
\begin{split}
q_1 & = \frac{1}{\sqrt{2}} \ket{0} + \frac{1}{\sqrt{2}} \ket{1} \\
    q_2 & = \ket{0}
\end{split}
\end{equation*}

\subsection{Measurement}
Measurement of a qubit $q$ with respect to the basis $\{\ket{0}, \ket{1}\}$ causes the state of $q$ to change to exactly one of $\ket{0}, \ket{1}$. In particular, if $q$ is in the state: 
\begin{equation*}
    q = \alpha \ket{0} + \beta \ket{1} 
\end{equation*}
then upon measurement, $q$ will collapse into state $\ket{0}$ with probability $|\alpha|^2$ and state $\ket{1}$ with probability $|\beta|^2$ \cite{rieffel2011quantum}.

\subsection{Entanglement}
The state of a two-qubit system $q_0, q_1$ is a vector in the tensor product of their respective state vector spaces \cite{rieffel2011quantum}. Therefore, if we use $\ket{0}, \ket{1}$ as the basis for each of the state vector spaces of $q_0, q_1$ respectively, then the basis for the state vector space of the two-qubit system is $\{ \ket{0}\otimes \ket{0}, \ket{0}\otimes \ket{1}, \ket{1}\otimes \ket{0}, \ket{1}\otimes\ket{1} \}$. \RTwoTwo{If two qubits $q_0, q_1$ have a state that cannot be represented as a complex scalar time basis vector, they are said to be entangled. For example, an example of an unentangled two-qubit state can be presented as follows,} 
\begin{equation*}
    -i\ket{1}\otimes \ket{0}
\end{equation*}
and an example of an entangled two-qubit state is presented as follows, 
\begin{equation*}
    \frac{1}{\sqrt{2}} \ket{0}\otimes\ket{0} - \frac{1}{\sqrt{2}} \ket{1}\otimes\ket{1} 
\end{equation*} 
Likewise, the state of an $n$-qubit system is a vector in the tensor product of their respective state vector spaces. For ease of reading, it is convention to represent the tensor product of basis vectors in the following form: 
\begin{equation*}
    \ket{b_1}\otimes \cdot\cdot\cdot \otimes \ket{b_n} = \ket{b_1...b_n} 
\end{equation*}
where $\ket{b_i}$ is a basis vector in the state vector space of qubit $q_i$. 

\subsection{Adiabatic Quantum Computing}
\Rone{In the previous sections, we introduced fundamental quantum primaries. In this section, we present how quantum annealing works in D-wave. }
The quantum state $\ket{\psi(t)}$ of a system of qubits varies with time according to the Schr\"odinger equation: 
\begin{equation*}
    i \frac{\partial}{\partial t} \ket{\psi(t)} = H(t) \ket{\psi(t)} 
\end{equation*}
where $H(t)$ is an operator called the Hamiltonian of the quantum system. At each time $t$, $H(t)$ describes the energy of the quantum state. Encoding the QUBO problem in a Hamiltonian $H_{QUBO}(t)$ means that the answer is the state of the qubits when $H_{QUBO}(t)$ is in the lowest energy state. \Rone{Since we do not know the lowest energy state of the QUBO prior, we could not initialize the $H_{QUBO}$ in that state}. Many researchers use the strategy of Adiabatic Quantum Computing to compute the lowest energy state of a complex quantum system, which is where we initialize the quantum system in the lowest energy state of some easy-to-compute Hamiltonian $H_{BASIC}$, and form our Hamiltonian as the interpolation between them: 
\begin{equation*}
    H(t) = (1-t/T)H_{BASIC}(t) + (t/T)H_{QUBO}(t) 
\end{equation*}
The Adiabatic Theorem states that as long as $0 \leq t \leq T$ is varied slowly enough, $H$ will remain in its ground state throughout the interval $[0, T]$, so when we measure the quantum state at time $T$, the system will be in its ground state and satisfy the QUBO problem \cite{farhi2000quantum}. 

\Rone{In the D-wave quantum computer, these processes are produced by the following ideas. First, the system begins with qubits, each in a superposition of 0 and 1, initially uncoupled. During quantum annealing, couplers and qubit biases are introduced, leading to entanglement among the qubits. This entangled state represents a multitude of possible solutions. By the end of the annealing process, each qubit settles into a classical state corresponding to the problem's minimum energy configuration or something very close. In D-Wave quantum computers, this entire process occurs within microseconds.}

\section{\RTwoTwo{Problem Definition and Motivations}}
\subsection{\RTwoTwo{Deep Visual Clustering}}

\RTwoThree{
Let $\mathcal{D}$ be a set of $N$ data points to be clustered. We define an embedding function $\mathcal{H}: \mathcal{D} \to \mathbb{R}^d$ that maps each data point $x_i \in \mathcal{D}$ to a latent space $\mathcal{X} \subset \mathbb{R}^{d}$. That is, each sample is transformed as:  
\begin{equation}
\mathbf{x}_i = \mathcal{H}(x_i), \quad \forall x_i \in \mathcal{D}.    
\end{equation}
}

\RTwoThree{
Let $\hat{y}_i$ be the ground-truth cluster ID assigned to the data point $x_i$. Since there are at most $N$ unique clusters, each $\hat{y}_i$ is an integer in the range $[0, N-1]$, forming a vector $\hat{Y} = (\hat{y}_1, \hat{y}_2, ..., \hat{y}_N) \in [0, N-1]^N$. A deep clustering algorithm $\Phi$ is then defined as a function that assigns a cluster label $y_i$ to each embedded data point $\mathbf{x}_i$, maximizing a clustering metric $\sigma(y_i, \hat{y}_i)$.
}
\RTwoThree{
Since the ground-truth labels $\hat{y}_i$ are inaccessible during training, the number of clusters and samples per cluster are unknown beforehand. To address this challenge, a \textit{divide-and-conquer} strategy is employed, decomposing $\Phi$ into three sub-functions: \textbf{Pre-processing} $\mathcal{K}$: Applies unsupervised clustering at a local level.  \textbf{In-processing} $\mathcal{M}$: Learns feature-based clustering representations. \textbf{Post-processing} $\mathcal{P}$: Merges clusters based on structural properties.  
}

\RTwoThree{
Thus, for each data point $\mathbf{x}_i$, the clustering function $\Phi$ is defined as: 
\begin{equation} \label{eqn:ClusteringAlgorithm}
\Phi(\mathbf{x}_i) = \mathcal{P}(\mathcal{M}(\mathcal{K}(\mathbf{x}_i, k))),
\end{equation}
where $\mathcal{K}(\mathbf{x}_i, k)$ retrieves $k$ nearest neighbors to form an initial local grouping. $\mathcal{M}$ refines the grouping through a learnable clustering process, and $\mathcal{P}$ further refines clusters via rule-based merging.
}

\RTwoThree{
Since $\mathcal{P}$ is typically a heuristic, non-learnable function \cite{yang2019learning,yang2020learning,Shen_2021_CVPR}, optimization focuses on $\mathcal{M}$, leading to the objective:  
\begin{equation}
    \label{eq:clustering}
\theta^*_{\mathcal{M}} = \arg \min_{\theta_{\mathcal{M}}} \mathbb{E}_{\mathbf{x}_i \sim p(\mathcal{X})} \left[ \mathcal{L} (\mathcal{M}(\mathcal{K}(\mathbf{x}_i, k)), \hat{q}_i) \right],
\end{equation}
where $\hat{q}_i$ represents the target clustering structure, and $\mathcal{L}$ is a suitable clustering or classification loss function.
}

\subsection{\RTwoTwo{Motivations}}
\begin{figure}
    \centering
    \includegraphics[width=\textwidth]{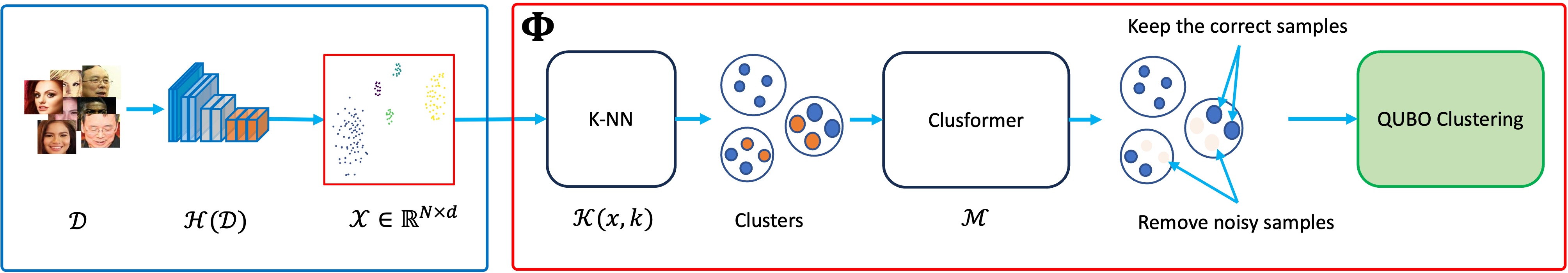}
    \caption{\RTwoTwo{Overall system architecture. The blue box indicates the deep visual feature extraction step. Specifically, $\mathcal{D}$ is the set of N data points. $\mathcal{H}$ is the feature extraction module and $\mathcal{X}$ is the corresponding feature set of the datapoint $\mathcal{D}$. The red box is the deep unsupervised algorithm where $\mathcal{K}$ denotes the K-NN algorithm, $\mathcal{M}$ is the deep cluster algorithm, i.e., Clusformer and $\mathcal{P}$ indicates the post-unsupervised clustering step. The green box indicates our focus in this paper.}}
    \label{fig:architecture}
\end{figure}

Given that the clustering method typically consists of three components, namely $\mathcal{K}$, $\mathcal{M}$, and $\mathcal{P}$, previous studies primarily concentrated on enhancing the performance of $\mathcal{M}$. However, the $\mathcal{P}$ also holds an important role contributing to the overall performance \cite{nguyen2021clusformer}. In most studies \cite{nguyen2021clusformer,mansoori2011frbc,ouyang2019rule}, the design of $\mathcal{P}$ is typically addressed with a simple rule-based algorithm, which often proves to be sensitive and reliant on the threshold chosen. Consequently, this sensitivity significantly impacts the overall performance of the clustering system, rendering it unstable. In this paper, we thoroughly investigate the design of $\mathcal{P}$ from an optimization perspective. To address this challenge, we propose a novel approach that leverages quantum machine techniques, improving unsupervised clustering performance. The overall architecture is illustrated in Fig \ref{fig:architecture}.

\section{Our Proposed Method - Quadratic Unconstrained
Binary Optimization Problem Approach}

In this section, we formulate the $\mathcal{P}$ component from the optimization perspective. Given a set of clusters $\mathcal{G} = \{g_0, g_1, \dots, g_{u-1}\}$. It is noted that $g_i = \mathcal{M}(\mathcal{K}(\mathbf{x}_i, k)$) is processed by deep network $\mathcal{M}$ to remove the noisy samples, thus, number of samples in each cluster $\vert g_i \vert$ can vary among clusters. \RTwoThree{In short, $g_i$ is generated and corresponding to $x_i$. In particular, given $x_i$, we first apply $\mathcal{K}$ (K-NN algorithm) to obtain $k$ nearest neighbors to form a cluster where $x_i$ holds the role as centroid. However, these clusters might have incorrect samples. Therefore, we apply Clusformer (denoted as $\mathcal{M}$) to detect and clear these samples. Finally, we obtain $g_i=\mathcal{M}(\mathcal{K}(x_i,k))$ is a cleaned cluster where $x_i$ is a centroid}. The objective of $\mathcal{P}$ can be defined as in Eqn. \eqref{eq:p_objective}.
\begin{equation}
    \label{eq:p_objective}
    \mathbf{max} \sum_{i,j}^{u} \mathbbm{1}_{c_i = c_j}
\end{equation}
where $\mathbbm{1}$ is the indicator function, $c_i$ is the \textbf{cluster id} of cluster $g_i$. Typically, the $c_i$ can be determined by the centroid $\mathbf{x}_i$. \RTwoTwo{This information is available in the training phase, but not in the testing phase}. For this reason,  $\mathcal{P}$ has to determine which clusters share the same ID. To address this problem, we find the optimal solution of Eqn. \eqref{eq:p_objective} using the Quadratic Unconstrained Binary Optimization Problem (QUBO). 

We can divide the $\mathcal{G}$ into two sets of clusters, denoted as $\mathcal{A}$ and $\mathcal{B}$. Let $\mathcal{A} = \{\textbf{a}_1, \textbf{a}_2, \dots \textbf{a}_r\}$ be a set of anchor clusters where each individual cluster $\textbf{a}_i$ is assigned a distinct ID. Let $\mathcal{B} = \{\textbf{b}_1, \textbf{b}_2, \dots \textbf{b}_s\}$ be the set of unknown clusters where we do not know their IDs. Our goal is to assign $\textbf{b}_j$ to a cluster $\textbf{a}_i$ if they share the same ID or add $\textbf{b}_j$ to be a new element of $\mathcal{A}$ if it does not match with any cluster inside $\mathcal{A}$. Formally, let $\mathbf{u} \in \mathbbm{R}^{r \times s}$ be the \textit{binary} matrix where \RTwoTwo{r and s are the number of clusters of $\mathcal{A}$ and $\mathcal{B}$. Each element $\textbf{u}_{ij} \in \textbf{u}$ either 0 or 1. Specially,}. 
\begin{align}
\textbf{u}_{ij} = 
\begin{cases}
  1  & \textbf{b}_j \text{ matches to } \textbf{a}_i \\
  0  & \text{ otherwise }
\end{cases} \nonumber
\end{align}
\RTwoTwo{To find the matrix $\mathbf{u}$, the formulation of the unsupervised problem includes following constraints.}
\subsection{\RTwoTwo{Cluster Constraint}}
Since $\textbf{b}_j$ could not match more than one $\textbf{a}_i$, \RTwoTwo{this constraint is represented} in Eqn. \eqref{eq:contrain_0}.
\begin{equation}
    \label{eq:contrain_0}
    0 \leq \sum_{i=1}^{r} \textbf{u}_{ij} \leq 1
\end{equation}
\RTwoTwo{In order to address this constraint, we propose the following optimization function}
\begin{align}
    \label{eq:qubo_3}
    f_0 = \sum_{j=1}^{s} \sum_{i=1}^{r-1}  \sum_{k=i+1}^{r} \textbf{u}_{ij} \textbf{u}_{kj}
\end{align}

\subsection{\RTwoTwo{Cluster Assignment}}
\RTwoTwo{The cluster $\textbf{b}_j$ is assigned to anchor $\textbf{a}_i$ if only if the cosine similarity distance, denoted as $d_{ij}$, is smallest. Since $\textbf{a}_i$ and $\textbf{b}_j$ are the cluster, we denote $\text{mean}(\textbf{a}_i)$ is the center vector of the cluster $\textbf{a}_i$. Specifically, we measure the average of feature vectors in the cluster as the center. Therefore, the distance $d_{ij}$ can be formulated as $d_{ij} = \text{cosine}(\text{mean}(\textbf{b}_j), \text{mean}(\textbf{a}_i))$. For that reason, we define the following optimization function as in the Eqn \eqref{eq:qubo_1}}:
\begin{align}
    \label{eq:qubo_1}
    f_1 = \sum_{i=1}^{r}\sum_{j=1}^{s} \textbf{u}_{ij} d_{ij}
\end{align}

\noindent
\subsection{\RTwoTwo{Triplet Assignment Constraint}}
\RTwoTwo{In clustering, the assignment of data points to clusters often depends not just on individual similarities as in the Eqn \eqref{eq:qubo_1} but also on pairwise relationships between data points. Let $\textbf{b}_j$ and $\textbf{b}_k$ are two separate clusters. They can be assigned to the same anchor $\textbf{a}_i$ by minimizing the function}
\begin{align}
    \label{eq:qubo_2}
    f_2 = \sum_{i=1}^{r}\sum_{j=1}^{s}\sum_{k=1}^{s}\textbf{u}_{ij} d_{ijk} \textbf{u}_{ik}
\end{align}
where $d_{ijk}$ is the triple costs among $\textbf{b}_j$, $\textbf{b}_k$ and $\textbf{a}_i$. It is measured by taking the average cost between each pair. That is $d_{ijk} = (d_{ij} + d_{ik} + d_{jk})/3$. 
\RTwoTwo{This equation accounts for the pairwise dependencies, ensuring that similar data points are grouped in the same cluster. In addition, $f_1$ is inherently quadratic due to the product of binary variables $\textbf{u}_{ij}$ and $\textbf{u}_{jk}$, which is key for encoding problems into QUBO. It makes $f_1$ directly suitable for optimization using quantum annealers or other solvers designed for QUBO problems. By minimizing $f_1$, the model encourages data points with a strong relationship (low $d_{ijk}$) to be assigned to the same cluster anchor $\textbf{a}_i$. It enhances intra-cluster similarity and improves clustering quality.
}

\subsection{\RTwoTwo{Overall Optimization Function}} 
In conclusion, the overall objective function can be presented as follows:
\begin{align}
    \label{eq:qubo_all}
    f &= f_0 + f_1 + f_2 \\ \nonumber
      &= 
    \sum_{j=1}^{s} \sum_{i=1}^{r-1}  \sum_{k=i+1}^{r} \textbf{u}_{ij} \textbf{u}_{kj} \\
      &+ \sum_{i=1}^{r}\sum_{j=1}^{s} \textbf{u}_{ij} d_{ij} + \sum_{i=1}^{r}\sum_{j=1}^{s}\sum_{k=1}^{s}\textbf{u}_{ij} d_{ijk} \textbf{u}_{ik} \\ \nonumber 
\end{align}

\subsection{\RTwoTwo{Pseudo Algorithm}}
\RTwoTwo{We have formulated the clustering component $\mathcal{P}$ as in the QUBO form. The details process of $\mathcal{P}$ is represented in the Algorithm \ref{algo:qubo_clustering}. In particular, the algorithm demonstrates steps to construct the anchor clusters $\mathcal{A}$ from a set of clusters $\mathcal{G}$. Firstly, we random sampling $b$ clusters from $\mathcal{G}$. Next, we solve the QUBO to find the assignments. Finally, we extend the anchor cluster $\textbf{a} \in \mathcal{A}$ by the new assignments.}

\begin{algorithm}[!ht]
\centering
\small
\caption{QUBO-based Clustering Algorithm}

\begin{algorithmic}
\STATE {\bfseries Input:}  
\quad $\mathcal{G} = \{g_0, g_1, \dots, g_{u-1}\}$: Set of elements to cluster  
\quad $b$: Batch size for processing  
\STATE {\bfseries Output:} List of anchor clusters $\mathcal{A}$  
\STATE Initialize an empty set of anchor clusters: $\mathcal{A} \gets \{\}$  

\WHILE{$|\mathcal{G}| > 0$}
    \STATE Select a random subset (batch) of elements: $\mathcal{B} \gets \textsf{SelectBatch}(\mathcal{G}, b)$  
    \STATE Solve the QUBO problem for batch clustering: $\mathbf{u} \gets \textsf{SolveQUBO}(\mathcal{A}, \mathcal{B})$  
    \STATE Update anchor clusters with the new assignments: $\mathcal{A} \gets \textsf{UpdateClusters}(\mathcal{A}, \mathbf{u})$  

    \STATE Remove processed batch $\mathcal{B}$ from $\mathcal{G}$: $\mathcal{G} \gets \textsf{RemoveBatch}(\mathcal{G}, \mathcal{B})$ 
\ENDWHILE  

\STATE \textbf{return} $\mathcal{A}$  

\end{algorithmic}
\label{algo:qubo_clustering}
\end{algorithm}

\section{Experiments and Results}
\subsection{Experiment Settings}
\noindent \textbf{Quantum Machine} AQC experiments are performed on a D-wave Advantage. The system contains at least 5000 qubits and 35,000 couplers implemented as superconducting qubits and Josephson junctions, respectively. Every qubit of the D-wave Advantage is connected to 15 other qubits, which needs to be reflected in the sparsity pattern of the cost matrix. If a denser matrix is required, chains of qubits are formed that represent a single state. The actual parameters
can vary due to defective qubits and couplers. All experiments use an annealing time of 1600 µs and an additional delay between measurements to reduce the intersample correlation. In the following, a single measurement combines an annealing cycle and the subsequent measurement. 

\noindent \textbf{Datasets}. This work is experimented on two datasets MNIST (Modified National Institute of Standards and Technology) \cite{deng2012mnist} and \RTwoTwo{CIFAR100} \cite{Krizhevsky09learningmultiple}. MNIST is a database of handwritten digits commonly used as a benchmark dataset for image classification tasks. It contains a collection of 60,000 training images and 10,000 testing images. The images are grayscale and have a resolution of 28x28 pixels. Each image represents a handwritten digit from 0 to 9, labeled accordingly. \RTwoTwo{CIFAR100} is a more challenging dataset compared to MNIST. It comprises 60,000 color images, equally divided into 50,000 training and 10,000 testing images. The images have a resolution of $32 \times 32$ pixels and are divided into 100 different classes. Each class contains 500 training images and 100 testing images.

\noindent \textbf{Deep Clustering Model}. In our work, we have employed AlexNet \cite{krizhevsky2017imagenet} as our feature generator $\mathcal{M}$. AlexNet is a convolutional neural network architecture that gained significant attention and revolutionized the field of computer vision when it won the ImageNet Large Scale Visual Recognition Challenge in 2012. It consists of multiple layers, including convolutional, pooling, and fully connected layers. AlexNet is known for its deep architecture and rectified linear units (ReLU) as activation functions, which help capture complex features from input images. Using AlexNet as our feature generator, we leverage its ability to extract high-level and discriminative features from images. The feature dimension $d = 64$. These features are then fed into the initial clustering algorithms $\mathcal{K}$. As defined in the previous works, we employ the K-Nearish Neighbors (K-NN) algorithm as $\mathcal{K}$ to find $k=32$ closest neighbors of a datapoint and form it as a cluster. It is important to note that these clusters contain false positive samples. For this reason, we employ Clusformer \cite{Nguyen_2021_CVPR} \RTwoThree{(as the role of  $\mathcal{M}$ in Eqn \eqref{eqn:ClusteringAlgorithm}) which is a transformer-based architecture designed for unsupervised large-scale recognition such as facial or visual landmark clustering. The goal of Clusformer is to detect noisy or hard samples inside clusters. This approach promises to refine the clustering results from K-NN.}

\noindent \textbf{Evaluation Protocol}
For face clustering, to measure the similarity between two clusters with a set of points, we use Fowlkes Mallows Score (FMS). This score is computed by taking the geometry mean of precision and recall. Thus, FMS $F_B$ is also called Pairwise-Fscore as follows,
\begin{equation}
    F_P = \frac{TP}{\sqrt{(TP + FP) \times (TP + FN)}}
\end{equation} 
where $TP$ is the number of point pairs in the same cluster in both ground truth and prediction. $FP$ is the number of point pairs in the same cluster in ground truth but not prediction. $FN$ is the number of point pairs in the same cluster in prediction but not in ground truth. Besides Pairwise F-score, BCubed-Fscore denoted as $F_B$ is also used for evaluation.

\subsection{Experimental Results} 

\noindent \textbf{k-Means}.
$k$-Means is a clustering algorithm that collects vectors into $k$ clusters such that the sum of the squared distances of each vector to the centroid of its cluster is minimized. In our experiments, the feature vectors of the images generated by the AlexNet are fed into the $k$-Means algorithm and sorted into $k$ clusters, where we vary $k$ to be both above and below the number of classes of the images. 
\cite{hartigan1979algorithm}

\noindent \textbf{HAC} 
Hierarchical Agglomerative Clustering (HAC) begins with each vector in a singleton set and takes in many clusters, and iteratively joins the two sets that minimize a linkage distance function until the number of sets is equal to the number of clusters. In the software implementation of this algorithm, we take an additional parameter: the number of neighbors. It is because comparing the linkage distance function between every set is computationally infeasible for large datasets. Therefore, the data is first organized into a $k$ nearest-neighbors graph where $k$ equals the number of neighbors. Then, two sets can only be joined if they are adjacent in this graph. 
\cite{nielsen2016clustering}

\noindent \textbf{DBSCAN}
Density-Based Spatial Clustering of Applications with Noise (DBSCAN) formalizes an intuitive idea of what a cluster is based on density by taking in a set of points and two parameters, $\epsilon$ and minimum number of points, which we will refer to as minpts. Ester et al., the creators of the algorithm, then define a cluster using the ideas of "direct density reachability" and "density connectivity." The DBSCAN algorithm operates by iterating over each point in its input set and checking if it has been classified. If it has not, it is considered a seed point for a cluster and grows the cluster according to the cluster definition provided by Ester et al. 
\cite{ester1996density}

\begin{table}
\small
\centering
\caption{Comparison in face clustering performance on unlabeled images in MNIST Database of our approach against prior SOTA approaches.  }
    \begin{tabular}{|l|l|cc|l|}
    \hline
    Method $/$ Metrics & Parameters & $F_P$ & $F_B$ & $NMI$ \\
    \hline 
    \multirow{5}{2em}{$k$-Means~\cite{lloyd1982least,sculley2010web}} 
    & k = 1 & 18.22 & 18.23 & 0 \\ 
    & k = 2 & 27.57 & 30.09 & 33.02 \\ 
    & k = 5 & 56.65 & 59.57 & 68.16 \\ 
    & {k = 10} & {92.26} & {92.40} & {90.56} \\ 
    & k = 15 & 80.75 & 80.50 & 85.28 \\ 

    \hline 
    \multirow{4}{2em}{HAC~\cite{sibson1973slink}} 
    & \# of clusters = 2, \;\;\;\# of neighbors = 100 
    & 31.62 & 32.67 & 41.11 \\ 
    & \# of clusters = 5, \;\;\;\# of neighbors = 100 
    & 50.98 & 64.09 & 70.39 \\ 
    & \# of clusters = 10, \;\# of neighbors = 100
    & 93.57 & 93.57 & 91.88 \\ 
    & \# of clusters = 15, \;\# of neighbors = 100
    & 85.61 & 85.43 & 87.45 \\ 

    \hline 
    \multirow{6}{2em}{DBSCAN~\cite{ester1996density}} 
    & $\epsilon$ = 13.0, \;\;\;\; minpts = 200, 
    & 47.41 & 65.28 & 68.01 \\ 
    & $\epsilon$ = 13.5, \;\;\;\; minpts = 200, 
    & 53.78 & 68.95 & 71.78 \\ 
    & $\epsilon$ = 14.0, \;\;\;\; minpts = 200, 
    & 60.04 & 72.47 & 75.04 \\ 
    & $\epsilon$ = 14.2, \;\; minpts = 200, 
    & 61.85 & 73.53 & 76.04 \\ 
    & $\epsilon$ = 14.5, \;\;\;\; minpts = 200, 
    & 54.53 & 68.45 & 71.59 \\ 
    & $\epsilon$ = 15.0, \;\; minpts = 200, 
    & 49.84 & 65.14 & 68.40 \\ 
    \hline
    \textbf{Clusformer \cite{Nguyen_2021_CVPR}} & - & {94.82} & {95.01} & {92.38} \\
    \textbf{Clusformer \cite{Nguyen_2021_CVPR} + QUBO} & - & \textbf{96.20} & \textbf{96.75} & \textbf{94.73} \\
    \hline 
    \end{tabular}
\label{tab:mnist}
\end{table}

\begin{table}
\small
\centering
\caption{Comparison in face clustering performance on unlabeled images in CIFAR100 Database of our approach against prior approaches.  }
    \begin{tabular}{|l|l|cc|l|}
    \hline
    Method $/$ Metrics & Parameters & $F_P$ & $F_B$ & $NMI$ \\
    \hline 
    \multirow{5}{2em}{$k$-Means~\cite{lloyd1982least,sculley2010web}} 
    & k = 1 & 18.16 & 18.18 & 0 \\ 
    & k = 2 & 26.43 & 29.24 & 30.01 \\ 
    & k = 5 & 38.44 & 48.99 & 49.44 \\ 
    & {k = 10} & {52.43} & {55.57} & {57.87} \\ 
    & k = 15 & 47.85 & 49.30 & 57.18 \\ 

    \hline 
    \multirow{4}{2em}{HAC~\cite{sibson1973slink}} 
    & \# of clusters = 2, \;\;\;\# of neighbors = 100 
    & 29.30 & 29.79 & 31.98 \\ 
    & \# of clusters = 5, \;\;\;\# of neighbors = 100 
    & 35.55 & 46.36 & 46.25 \\ 
    & \# of clusters = 10, \;\# of neighbors = 100
    & 48.07 & 51.80 & 54.08 \\ 
    & \# of clusters = 15, \;\# of neighbors = 100
    & 42.55 & 46.65 & 53.05 \\ 

    \hline 
    \multirow{6}{2em}{DBSCAN~\cite{ester1996density}} 
    & $\epsilon$ = 0.8, \;\;\;\; minpts = 950, 
    & 21.33 & 25.93 & 18.40 \\ 
    & $\epsilon$ {= 0.84}, \;\; {minpts = 950}, 
    & {27.36} & 28.52 & 28.15 \\ 
    & $\epsilon$ = 0.9, \;\;\;\; minpts = 950, 
    & 18.14 & 19.03 & 1.71 \\ 
    & $\epsilon$ = 0.8, \;\;\;\; minpts = 1000, 
    & 19.51 & 24.75 & 13.68 \\ 
    & $\epsilon$ {= 0.84}, \;\; {minpts = 1000}, 
    & 27.30 & {28.56} & {28.19} \\ 
    & $\epsilon$ = 0.9, \;\;\;\; minpts = 1000, 
    & 18.14 & 19.08 & 1.84 \\ 
    \hline
    \textbf{Clusformer \cite{Nguyen_2021_CVPR}} & - & {54.36} & {57.28} & {58.38} \\
    \textbf{Clusformer \cite{Nguyen_2021_CVPR} + QUBO} & - & \textbf{56.98} & \textbf{59.32} & \textbf{60.10} \\
    \hline 
    \end{tabular}
\label{tab:CIFAR100}
\end{table}

In order to determine the optimal parameters for achieving superior performance, we conducted a series of experiments that involved exploring a wide range of settings for the K-Mean, HAC, and DBSCAN algorithms. Specifically, we experimented with different numbers of clusters, ranging from 1 to 15, for both K-Mean and HAC. For DBSCAN, we varied the settings, such as $\epsilon$ and $minpts$, to find the most suitable values. Our investigation revealed that K-Mean and HAC algorithms excel when the exact number of clusters, denoted as $k=10$, is known in advance. However, it is essential to note that this information is often unavailable in practical scenarios, making the task more challenging.

To evaluate the effectiveness of the algorithms, we conducted clustering experiments on the CIFAR100 database and recorded the results in Table \ref{tab:mnist}. The table provides a comprehensive overview of the clustering performance of K-Mean, HAC, and DBSCAN, allowing us to make meaningful comparisons. Interestingly, the Clusformer algorithm emerged as the standout performer, outperforming the previous methods by a substantial margin. Its superiority is evident across multiple performance metrics. For instance, Clusformer achieved an impressive 94.82\% for $F_P$ (precision), 95.01\% for $F_B$ (recall), and 92.38\% for $NMI$, surpassing the results obtained by K-Mean, HAC, and DBSCAN by approximately 1.25\% to 2.56\%.

Moreover, we explored the application of Quantum Computing to enhance the performance of Clusformer. By leveraging the power of Quantum Computers to solve the Quadratic Unconstrained Binary Optimization (QUBO) problem, we achieved even more significant performance improvements. The utilization of QUBO resulted in Clusformer achieving remarkable scores of 96.20\% for $F_P$, 96.75\% for $F_B$, and 94.73\% for $NMI$. These results solidify Clusformer's superiority over previous methods and further extend the improvement gap.

Similarly, We conduct the same experiment on the CIFAR database as shown in Table \ref{tab:CIFAR100}. We obtained a similar conclusion as in the CIFAR100 database. It demonstrates that leveraging Quantum Machine to solve QUBO can further benefit the unsupervised clustering problem.

\section{Conclusions} 
In this work, we have formulated a QUBO clustering algorithm for visual clustering and solved it efficiently using adiabatic quantum computing. We used experiments on the \RTwoTwo{CIFAR100} and MNIST datasets to demonstrate the performance of our QUBO algorithm. Considering current limitations inherent to the hardware of quantum computers, there is tremendous potential in the future to solve more complex and descriptive QUBO formulations using adiabatic quantum computing. 

\section{Availability of data and materials}
The data that supports the findings of this study can be found at \href{https://www.cs.toronto.edu/~kriz/cifar.html}{https://www.cs.toronto.edu/~kriz/cifar.html} and \href{http://yann.lecun.com/exdb/mnist/}{http://yann.lecun.com/exdb/mnist/}

\newpage

\begin{center}
    \section*{Author Biography}
\end{center}

\textbf{Xuan-Bac Nguyen} is currently a Ph.D. student at the Department of Computer Science and Computer Engineering of the University of Arkansas. He received his M.Sc. degree in Computer Science from the Electrical and Computer Engineering Department at Chonnam National University, South Korea, in 2020.  He received his B.Sc. degree in Electronics and Telecommunications from the University of Engineering and Technology, VNU, in 2015. In 2016, he was a software engineer in Yokohama, Japan. His research interests include Quantum Machine Learning, Face Recognition, Facial Expression, and Medical Image Processing.

\textbf{Hugh Churchill} is a Professor and 21st Century Chair in Nanophysics in the Department of Physics at the University of Arkansas (UA).  He is also Associate Director for Operations of the MonArk NSF Quantum Foundry.  He received a Ph.D. in Physics from Harvard University in 2012 and was a Pappalardo postdoctoral fellow in Physics at MIT.  Since 2015 he has led the Quantum Device Laboratory at UA, pursuing research interests in quantum materials and devices, low-dimensional materials, quantum transport, optoelectronics, and automation of experiments.

\textbf{Khoa Luu} 
is an Assistant Professor and the Director of the Computer Vision and Image Understanding (CVIU) Lab in the Department of Electrical Engineering and Computer Science (EECS) at the University of Arkansas (UA), Fayetteville, US. He is affiliated with the NSF MonARK Quantum Foundry. He is an Associate Editor of the IEEE Access Journal and the Multimedia Tools and Applications Journal, Springer Nature. He is also the Area Chair in CVPR 2023, CVPR 2024, NeurIPS 2024, WACV 2025, and ICLR 2025. He was the Research Project Director at the Cylab Biometrics Center at Carnegie Mellon University (CMU), USA. His research interests focus on various topics, including Quantum Machine Learning, Biometrics, Smart Health, and Precision Agriculture. He has received eight patents and three Best Paper Awards and coauthored 120+ papers in conferences, technical reports, and journals. He was a Vice-Chair of the Montreal Chapter of the IEEE Systems, Man, and Cybernetics Society in Canada from September 2009 to March 2011. He was the Technical Program Chair at the IEEE GreenTech Conference 2024 and a co-organizer and chair of the CVPR Precognition Workshop in 2019-2024, MICCAI Workshop in 2019 and 2020, and ICCV Workshop in 2021. He was a PC member of AAAI, ICPRAI in 2020, 2021, 2022. He has been an active reviewer for several top-tier conferences and journals, such as CVPR, ICCV, ECCV, NeurIPS, ICLR, FG, BTAS, IEEE-TPAMI, IEEE-TIP, IEEE Access, Journal of Pattern Recognition, Journal of Image and Vision Computing, Journal of Signal Processing, and Journal of Intelligence Review.

\textbf{Samee U. Khan} received a Ph.D. in 2007 from the University of Texas, Arlington, TX. Currently, he is Professor of Electrical \& Computer Engineering at Mississippi State University (MSU) and served as Department Head from Aug. 2020 to July 2024. Before arriving at MSU, he was Cluster Lead (2016-2020) for Computer Systems Research at the National Science Foundation and the Walter B. Booth Professor at North Dakota State University. His research interests include optimization, robustness, and security of computer systems. His work has appeared in over 450 publications. He is the associate editor of IEEE Transactions on Cloud Computing and the Journal of Parallel and Distributed Computing. He is a Fellow of the IET and BCS. He is a Distinguished Member of the ACM and a Senior Member of the IEEE. 
He has won several awards, including the Best Paper Award (Systems Track), IEEE Cloud Summit, 2024; IEEE R3 Outstanding Engineer Award, 2024; IEEE Computer Society Distinguished Contributor Award, 2022 (inducted in the inaugural class); IEEE ComSoc Technical Committee on Big Data Best Journal Paper Award, 2019; IEEE-USA Professional Achievement Award, 2016; IEEE Golden Core Member Award, 2016; IEEE TCSC Award for Excellence in Scalable Computing Research (Middle Career Researcher), 2016; IEEE Computer Society Meritorious Service Certificate, 2016; Tapestry of Diverse Talents Award, North Dakota State University (NDSU), ND, USA, 2016; Exemplary Editor, IEEE Communications Surveys and Tutorials, IEEE Communications Society, 2014; Outstanding Summer Undergraduate Research Faculty Mentor Award, NDSU, ND, USA, 2013; Best Paper Award, IEEE Intl. Conf. on Scalable Computing and Communications (ScalCom), 2012; Sudhir Mehta Memorial International Faculty Award, NDSU, ND, USA, 2012; Best Paper Award, ACM/IEEE Intl. Conf. on Green Computing \& Communications (GreenCom), 2010.

\bibliography{sn-bibliography}%


\begin{thebibliography}{33}
\ifx \bisbn   \undefined \def \bisbn  #1{ISBN #1}\fi
\ifx \binits  \undefined \def \binits#1{#1}\fi
\ifx \bauthor  \undefined \def \bauthor#1{#1}\fi
\ifx \batitle  \undefined \def \batitle#1{#1}\fi
\ifx \bjtitle  \undefined \def \bjtitle#1{#1}\fi
\ifx \bvolume  \undefined \def \bvolume#1{\textbf{#1}}\fi
\ifx \byear  \undefined \def \byear#1{#1}\fi
\ifx \bissue  \undefined \def \bissue#1{#1}\fi
\ifx \bfpage  \undefined \def \bfpage#1{#1}\fi
\ifx \blpage  \undefined \def \blpage #1{#1}\fi
\ifx \burl  \undefined \def \burl#1{\textsf{#1}}\fi
\ifx \doiurl  \undefined \def \doiurl#1{\url{https://doi.org/#1}}\fi
\ifx \betal  \undefined \def \betal{\textit{et al.}}\fi
\ifx \binstitute  \undefined \def \binstitute#1{#1}\fi
\ifx \binstitutionaled  \undefined \def \binstitutionaled#1{#1}\fi
\ifx \bctitle  \undefined \def \bctitle#1{#1}\fi
\ifx \beditor  \undefined \def \beditor#1{#1}\fi
\ifx \bpublisher  \undefined \def \bpublisher#1{#1}\fi
\ifx \bbtitle  \undefined \def \bbtitle#1{#1}\fi
\ifx \bedition  \undefined \def \bedition#1{#1}\fi
\ifx \bseriesno  \undefined \def \bseriesno#1{#1}\fi
\ifx \blocation  \undefined \def \blocation#1{#1}\fi
\ifx \bsertitle  \undefined \def \bsertitle#1{#1}\fi
\ifx \bsnm \undefined \def \bsnm#1{#1}\fi
\ifx \bsuffix \undefined \def \bsuffix#1{#1}\fi
\ifx \bparticle \undefined \def \bparticle#1{#1}\fi
\ifx \barticle \undefined \def \barticle#1{#1}\fi
\bibcommenthead
\ifx \bconfdate \undefined \def \bconfdate #1{#1}\fi
\ifx \botherref \undefined \def \botherref #1{#1}\fi
\ifx \url \undefined \def \url#1{\textsf{#1}}\fi
\ifx \bchapter \undefined \def \bchapter#1{#1}\fi
\ifx \bbook \undefined \def \bbook#1{#1}\fi
\ifx \bcomment \undefined \def \bcomment#1{#1}\fi
\ifx \oauthor \undefined \def \oauthor#1{#1}\fi
\ifx \citeauthoryear \undefined \def \citeauthoryear#1{#1}\fi
\ifx \endbibitem  \undefined \def \endbibitem {}\fi
\ifx \bconflocation  \undefined \def \bconflocation#1{#1}\fi
\ifx \arxivurl  \undefined \def \arxivurl#1{\textsf{#1}}\fi
\csname PreBibitemsHook\endcsname

\bibitem[\protect\citeauthoryear{Yang et~al.}{2019}]{yang2019learning}
\begin{bchapter}
\bauthor{\bsnm{Yang}, \binits{L.}},
\bauthor{\bsnm{Zhan}, \binits{X.}},
\bauthor{\bsnm{Chen}, \binits{D.}},
\bauthor{\bsnm{Yan}, \binits{J.}},
\bauthor{\bsnm{Loy}, \binits{C.C.}},
\bauthor{\bsnm{Lin}, \binits{D.}}:
\bctitle{Learning to cluster faces on an affinity graph}.
In: \bbtitle{Proceedings of the IEEE Conference on Computer Vision and Pattern
  Recognition (CVPR)}
(\byear{2019})
\end{bchapter}
\endbibitem

\bibitem[\protect\citeauthoryear{Yang et~al.}{2020}]{yang2020learning}
\begin{bchapter}
\bauthor{\bsnm{Yang}, \binits{L.}},
\bauthor{\bsnm{Chen}, \binits{D.}},
\bauthor{\bsnm{Zhan}, \binits{X.}},
\bauthor{\bsnm{Zhao}, \binits{R.}},
\bauthor{\bsnm{Loy}, \binits{C.C.}},
\bauthor{\bsnm{Lin}, \binits{D.}}:
\bctitle{Learning to cluster faces via confidence and connectivity estimation}.
In: \bbtitle{Proceedings of the IEEE Conference on Computer Vision and Pattern
  Recognition}
(\byear{2020})
\end{bchapter}
\endbibitem

\bibitem[\protect\citeauthoryear{Nguyen et~al.}{2021}]{Nguyen_2021_CVPR}
\begin{bchapter}
\bauthor{\bsnm{Nguyen}, \binits{X.-B.}},
\bauthor{\bsnm{Bui}, \binits{D.T.}},
\bauthor{\bsnm{Duong}, \binits{C.N.}},
\bauthor{\bsnm{Bui}, \binits{T.D.}},
\bauthor{\bsnm{Luu}, \binits{K.}}:
\bctitle{Clusformer: A transformer based clustering approach to unsupervised
  large-scale face and visual landmark recognition}.
In: \bbtitle{Proceedings of the IEEE/CVF Conference on Computer Vision and
  Pattern Recognition (CVPR)},
pp. \bfpage{10847}--\blpage{10856}
(\byear{2021})
\end{bchapter}
\endbibitem

\bibitem[\protect\citeauthoryear{Kerenidis et~al.}{2019}]{kerenidis2019q}
\begin{botherref}
\oauthor{\bsnm{Kerenidis}, \binits{I.}},
\oauthor{\bsnm{Landman}, \binits{J.}},
\oauthor{\bsnm{Luongo}, \binits{A.}},
\oauthor{\bsnm{Prakash}, \binits{A.}}:
q-means: A quantum algorithm for unsupervised machine learning.
Advances in neural information processing systems
\textbf{32}
(2019)
\end{botherref}
\endbibitem

\bibitem[\protect\citeauthoryear{Cerezo et~al.}{2022}]{cerezo2022challenges}
\begin{barticle}
\bauthor{\bsnm{Cerezo}, \binits{M.}},
\bauthor{\bsnm{Verdon}, \binits{G.}},
\bauthor{\bsnm{Huang}, \binits{H.-Y.}},
\bauthor{\bsnm{Cincio}, \binits{L.}},
\bauthor{\bsnm{Coles}, \binits{P.J.}}:
\batitle{Challenges and opportunities in quantum machine learning}.
\bjtitle{Nature Computational Science}
\bvolume{2}(\bissue{9}),
\bfpage{567}--\blpage{576}
(\byear{2022})
\end{barticle}
\endbibitem

\bibitem[\protect\citeauthoryear{Nguyen et~al.}{2024}]{nguyen2024theory}
\begin{barticle}
\bauthor{\bsnm{Nguyen}, \binits{Q.T.}},
\bauthor{\bsnm{Schatzki}, \binits{L.}},
\bauthor{\bsnm{Braccia}, \binits{P.}},
\bauthor{\bsnm{Ragone}, \binits{M.}},
\bauthor{\bsnm{Coles}, \binits{P.J.}},
\bauthor{\bsnm{Sauvage}, \binits{F.}},
\bauthor{\bsnm{Larocca}, \binits{M.}},
\bauthor{\bsnm{Cerezo}, \binits{M.}}:
\batitle{Theory for equivariant quantum neural networks}.
\bjtitle{PRX Quantum}
\bvolume{5}(\bissue{2}),
\bfpage{020328}
(\byear{2024})
\end{barticle}
\endbibitem

\bibitem[\protect\citeauthoryear{Schatzki
  et~al.}{2024}]{schatzki2024theoretical}
\begin{barticle}
\bauthor{\bsnm{Schatzki}, \binits{L.}},
\bauthor{\bsnm{Larocca}, \binits{M.}},
\bauthor{\bsnm{Nguyen}, \binits{Q.T.}},
\bauthor{\bsnm{Sauvage}, \binits{F.}},
\bauthor{\bsnm{Cerezo}, \binits{M.}}:
\batitle{Theoretical guarantees for permutation-equivariant quantum neural
  networks}.
\bjtitle{npj Quantum Information}
\bvolume{10}(\bissue{1}),
\bfpage{12}
(\byear{2024})
\end{barticle}
\endbibitem

\bibitem[\protect\citeauthoryear{Zaech et~al.}{2022}]{zaech2022adiabatic}
\begin{bchapter}
\bauthor{\bsnm{Zaech}, \binits{J.-N.}},
\bauthor{\bsnm{Liniger}, \binits{A.}},
\bauthor{\bsnm{Danelljan}, \binits{M.}},
\bauthor{\bsnm{Dai}, \binits{D.}},
\bauthor{\bsnm{Van~Gool}, \binits{L.}}:
\bctitle{Adiabatic quantum computing for multi object tracking}.
In: \bbtitle{Proceedings of the IEEE/CVF Conference on Computer Vision and
  Pattern Recognition},
pp. \bfpage{8811}--\blpage{8822}
(\byear{2022})
\end{bchapter}
\endbibitem

\bibitem[\protect\citeauthoryear{Doan et~al.}{2022}]{doan2022hybrid}
\begin{bchapter}
\bauthor{\bsnm{Doan}, \binits{A.-D.}},
\bauthor{\bsnm{Sasdelli}, \binits{M.}},
\bauthor{\bsnm{Suter}, \binits{D.}},
\bauthor{\bsnm{Chin}, \binits{T.-J.}}:
\bctitle{A hybrid quantum-classical algorithm for robust fitting}.
In: \bbtitle{Proceedings of the IEEE/CVF Conference on Computer Vision and
  Pattern Recognition},
pp. \bfpage{417}--\blpage{427}
(\byear{2022})
\end{bchapter}
\endbibitem

\bibitem[\protect\citeauthoryear{Arthur and Date}{2021}]{arthur2021balanced}
\begin{barticle}
\bauthor{\bsnm{Arthur}, \binits{D.}},
\bauthor{\bsnm{Date}, \binits{P.}}:
\batitle{Balanced k-means clustering on an adiabatic quantum computer}.
\bjtitle{Quantum Information Processing}
\bvolume{20},
\bfpage{1}--\blpage{30}
(\byear{2021})
\end{barticle}
\endbibitem

\bibitem[\protect\citeauthoryear{Ren et~al.}{2022}]{ren2022deep}
\begin{botherref}
\oauthor{\bsnm{Ren}, \binits{Y.}},
\oauthor{\bsnm{Pu}, \binits{J.}},
\oauthor{\bsnm{Yang}, \binits{Z.}},
\oauthor{\bsnm{Xu}, \binits{J.}},
\oauthor{\bsnm{Li}, \binits{G.}},
\oauthor{\bsnm{Pu}, \binits{X.}},
\oauthor{\bsnm{Yu}, \binits{P.S.}},
\oauthor{\bsnm{He}, \binits{L.}}:
Deep clustering: A comprehensive survey.
arXiv preprint arXiv:2210.04142
(2022)
\end{botherref}
\endbibitem

\bibitem[\protect\citeauthoryear{Lopes and Pedronette}{2023}]{Lopes_2023_WACV}
\begin{bchapter}
\bauthor{\bsnm{Lopes}, \binits{L.T.}},
\bauthor{\bsnm{Pedronette}, \binits{D.C.G.a.}}:
\bctitle{Self-supervised clustering based on manifold learning and graph
  convolutional networks}.
In: \bbtitle{Proceedings of the IEEE/CVF Winter Conference on Applications of
  Computer Vision (WACV)},
pp. \bfpage{5634}--\blpage{5643}
(\byear{2023})
\end{bchapter}
\endbibitem

\bibitem[\protect\citeauthoryear{Bo et~al.}{2020}]{bo2020structural}
\begin{bchapter}
\bauthor{\bsnm{Bo}, \binits{D.}},
\bauthor{\bsnm{Wang}, \binits{X.}},
\bauthor{\bsnm{Shi}, \binits{C.}},
\bauthor{\bsnm{Zhu}, \binits{M.}},
\bauthor{\bsnm{Lu}, \binits{E.}},
\bauthor{\bsnm{Cui}, \binits{P.}}:
\bctitle{Structural deep clustering network}.
In: \bbtitle{Proceedings of the Web Conference 2020},
pp. \bfpage{1400}--\blpage{1410}
(\byear{2020})
\end{bchapter}
\endbibitem

\bibitem[\protect\citeauthoryear{Ling et~al.}{2022}]{ling2022vision}
\begin{botherref}
\oauthor{\bsnm{Ling}, \binits{H.-B.}},
\oauthor{\bsnm{Zhu}, \binits{B.}},
\oauthor{\bsnm{Huang}, \binits{D.}},
\oauthor{\bsnm{Chen}, \binits{D.-H.}},
\oauthor{\bsnm{Wang}, \binits{C.-D.}},
\oauthor{\bsnm{Lai}, \binits{J.-H.}}:
Vision transformer for contrastive clustering.
arXiv preprint arXiv:2206.12925
(2022)
\end{botherref}
\endbibitem

\bibitem[\protect\citeauthoryear{Chen et~al.}{2022}]{chen2022graph}
\begin{barticle}
\bauthor{\bsnm{Chen}, \binits{J.}},
\bauthor{\bsnm{Han}, \binits{J.}},
\bauthor{\bsnm{Meng}, \binits{X.}},
\bauthor{\bsnm{Li}, \binits{Y.}},
\bauthor{\bsnm{Li}, \binits{H.}}:
\batitle{Graph convolutional network combined with semantic feature guidance
  for deep clustering}.
\bjtitle{Tsinghua Science and Technology}
\bvolume{27}(\bissue{5}),
\bfpage{855}--\blpage{868}
(\byear{2022})
\end{barticle}
\endbibitem

\bibitem[\protect\citeauthoryear{Kipf and Welling}{2016}]{kipf2016semi}
\begin{botherref}
\oauthor{\bsnm{Kipf}, \binits{T.N.}},
\oauthor{\bsnm{Welling}, \binits{M.}}:
Semi-supervised classification with graph convolutional networks.
arXiv preprint arXiv:1609.02907
(2016)
\end{botherref}
\endbibitem

\bibitem[\protect\citeauthoryear{Huo et~al.}{2021}]{huo2021caegcn}
\begin{botherref}
\oauthor{\bsnm{Huo}, \binits{G.}},
\oauthor{\bsnm{Zhang}, \binits{Y.}},
\oauthor{\bsnm{Gao}, \binits{J.}},
\oauthor{\bsnm{Wang}, \binits{B.}},
\oauthor{\bsnm{Hu}, \binits{Y.}},
\oauthor{\bsnm{Yin}, \binits{B.}}:
Caegcn: Cross-attention fusion based enhanced graph convolutional network for
  clustering.
IEEE Transactions on Knowledge and Data Engineering
(2021)
\end{botherref}
\endbibitem

\bibitem[\protect\citeauthoryear{Scarselli et~al.}{2009}]{4700287}
\begin{barticle}
\bauthor{\bsnm{Scarselli}, \binits{F.}},
\bauthor{\bsnm{Gori}, \binits{M.}},
\bauthor{\bsnm{Tsoi}, \binits{A.C.}},
\bauthor{\bsnm{Hagenbuchner}, \binits{M.}},
\bauthor{\bsnm{Monfardini}, \binits{G.}}:
\batitle{The graph neural network model}.
\bjtitle{IEEE Transactions on Neural Networks}
\bvolume{20}(\bissue{1}),
\bfpage{61}--\blpage{80}
(\byear{2009})
\doiurl{10.1109/TNN.2008.2005605}
\end{barticle}
\endbibitem

\bibitem[\protect\citeauthoryear{Nguyen et~al.}{2021}]{nguyen2021clusformer}
\begin{bchapter}
\bauthor{\bsnm{Nguyen}, \binits{X.-B.}},
\bauthor{\bsnm{Bui}, \binits{D.T.}},
\bauthor{\bsnm{Duong}, \binits{C.N.}},
\bauthor{\bsnm{Bui}, \binits{T.D.}},
\bauthor{\bsnm{Luu}, \binits{K.}}:
\bctitle{Clusformer: A transformer based clustering approach to unsupervised
  large-scale face and visual landmark recognition}.
In: \bbtitle{Proceedings of the IEEE/CVF Conference on Computer Vision and
  Pattern Recognition},
pp. \bfpage{10847}--\blpage{10856}
(\byear{2021})
\end{bchapter}
\endbibitem

\bibitem[\protect\citeauthoryear{Rieffel and Polak}{2011}]{rieffel2011quantum}
\begin{bbook}
\bauthor{\bsnm{Rieffel}, \binits{E.G.}},
\bauthor{\bsnm{Polak}, \binits{W.H.}}:
\bbtitle{Quantum Computing: A Gentle Introduction}.
\bpublisher{MIT Press}, \blocation{???}
(\byear{2011})
\end{bbook}
\endbibitem

\bibitem[\protect\citeauthoryear{Farhi et~al.}{2000}]{farhi2000quantum}
\begin{botherref}
\oauthor{\bsnm{Farhi}, \binits{E.}},
\oauthor{\bsnm{Goldstone}, \binits{J.}},
\oauthor{\bsnm{Gutmann}, \binits{S.}},
\oauthor{\bsnm{Sipser}, \binits{M.}}:
Quantum computation by adiabatic evolution.
arXiv preprint quant-ph/0001106
(2000)
\end{botherref}
\endbibitem

\bibitem[\protect\citeauthoryear{Shen et~al.}{2021}]{Shen_2021_CVPR}
\begin{bchapter}
\bauthor{\bsnm{Shen}, \binits{S.}},
\bauthor{\bsnm{Li}, \binits{W.}},
\bauthor{\bsnm{Zhu}, \binits{Z.}},
\bauthor{\bsnm{Huang}, \binits{G.}},
\bauthor{\bsnm{Du}, \binits{D.}},
\bauthor{\bsnm{Lu}, \binits{J.}},
\bauthor{\bsnm{Zhou}, \binits{J.}}:
\bctitle{Structure-aware face clustering on a large-scale graph with 107
  nodes}.
In: \bbtitle{Proceedings of the IEEE/CVF Conference on Computer Vision and
  Pattern Recognition (CVPR)},
pp. \bfpage{9085}--\blpage{9094}
(\byear{2021})
\end{bchapter}
\endbibitem

\bibitem[\protect\citeauthoryear{Mansoori}{2011}]{mansoori2011frbc}
\begin{barticle}
\bauthor{\bsnm{Mansoori}, \binits{E.G.}}:
\batitle{Frbc: A fuzzy rule-based clustering algorithm}.
\bjtitle{IEEE transactions on fuzzy systems}
\bvolume{19}(\bissue{5}),
\bfpage{960}--\blpage{971}
(\byear{2011})
\end{barticle}
\endbibitem

\bibitem[\protect\citeauthoryear{Ouyang et~al.}{2019}]{ouyang2019rule}
\begin{barticle}
\bauthor{\bsnm{Ouyang}, \binits{T.}},
\bauthor{\bsnm{Pedrycz}, \binits{W.}},
\bauthor{\bsnm{Pizzi}, \binits{N.J.}}:
\batitle{Rule-based modeling with dbscan-based information granules}.
\bjtitle{IEEE Transactions on Cybernetics}
\bvolume{51}(\bissue{7}),
\bfpage{3653}--\blpage{3663}
(\byear{2019})
\end{barticle}
\endbibitem

\bibitem[\protect\citeauthoryear{Deng}{2012}]{deng2012mnist}
\begin{barticle}
\bauthor{\bsnm{Deng}, \binits{L.}}:
\batitle{The mnist database of handwritten digit images for machine learning
  research [best of the web]}.
\bjtitle{IEEE signal processing magazine}
\bvolume{29}(\bissue{6}),
\bfpage{141}--\blpage{142}
(\byear{2012})
\end{barticle}
\endbibitem

\bibitem[\protect\citeauthoryear{Krizhevsky}{2009}]{Krizhevsky09learningmultiple}
\begin{botherref}
\oauthor{\bsnm{Krizhevsky}, \binits{A.}}:
Learning multiple layers of features from tiny images.
Technical report
(2009)
\end{botherref}
\endbibitem

\bibitem[\protect\citeauthoryear{Krizhevsky
  et~al.}{2017}]{krizhevsky2017imagenet}
\begin{barticle}
\bauthor{\bsnm{Krizhevsky}, \binits{A.}},
\bauthor{\bsnm{Sutskever}, \binits{I.}},
\bauthor{\bsnm{Hinton}, \binits{G.E.}}:
\batitle{Imagenet classification with deep convolutional neural networks}.
\bjtitle{Communications of the ACM}
\bvolume{60}(\bissue{6}),
\bfpage{84}--\blpage{90}
(\byear{2017})
\end{barticle}
\endbibitem

\bibitem[\protect\citeauthoryear{Hartigan and
  Wong}{1979}]{hartigan1979algorithm}
\begin{barticle}
\bauthor{\bsnm{Hartigan}, \binits{J.A.}},
\bauthor{\bsnm{Wong}, \binits{M.A.}}:
\batitle{Algorithm as 136: A k-means clustering algorithm}.
\bjtitle{Journal of the royal statistical society. series c (applied
  statistics)}
\bvolume{28}(\bissue{1}),
\bfpage{100}--\blpage{108}
(\byear{1979})
\end{barticle}
\endbibitem

\bibitem[\protect\citeauthoryear{Nielsen}{2016}]{nielsen2016clustering}
\begin{bbook}
\bauthor{\bsnm{Nielsen}, \binits{F.}}:
\bbtitle{Hierarchical Clustering},
pp. \bfpage{195}--\blpage{211}
(\byear{2016}).
\doiurl{10.1007/978-3-319-21903-5_8}
\end{bbook}
\endbibitem

\bibitem[\protect\citeauthoryear{Ester et~al.}{1996}]{ester1996density}
\begin{bchapter}
\bauthor{\bsnm{Ester}, \binits{M.}},
\bauthor{\bsnm{Kriegel}, \binits{H.-P.}},
\bauthor{\bsnm{Sander}, \binits{J.}},
\bauthor{\bsnm{Xu}, \binits{X.}}, \betal:
\bctitle{A density-based algorithm for discovering clusters in large spatial
  databases with noise.}
In: \bbtitle{Kdd},
vol. \bseriesno{96},
pp. \bfpage{226}--\blpage{231}
(\byear{1996})
\end{bchapter}
\endbibitem

\bibitem[\protect\citeauthoryear{Lloyd}{1982}]{lloyd1982least}
\begin{barticle}
\bauthor{\bsnm{Lloyd}, \binits{S.}}:
\batitle{Least squares quantization in pcm}.
\bjtitle{IEEE transactions on information theory}
\bvolume{28}(\bissue{2}),
\bfpage{129}--\blpage{137}
(\byear{1982})
\end{barticle}
\endbibitem

\bibitem[\protect\citeauthoryear{Sculley}{2010}]{sculley2010web}
\begin{bchapter}
\bauthor{\bsnm{Sculley}, \binits{D.}}:
\bctitle{Web-scale k-means clustering}.
In: \bbtitle{Proceedings of the 19th International Conference on World Wide
  Web},
pp. \bfpage{1177}--\blpage{1178}
(\byear{2010})
\end{bchapter}
\endbibitem

\bibitem[\protect\citeauthoryear{Sibson}{1973}]{sibson1973slink}
\begin{barticle}
\bauthor{\bsnm{Sibson}, \binits{R.}}:
\batitle{Slink: an optimally efficient algorithm for the single-link cluster
  method}.
\bjtitle{The computer journal}
\bvolume{16}(\bissue{1}),
\bfpage{30}--\blpage{34}
(\byear{1973})
\end{barticle}
\endbibitem

\end{thebibliography}

\end{document}